\documentclass[10pt,twoside]{hsqcd}
\usepackage{epsf,amsmath}
\usepackage{graphicx}
\def\lsim{\raise0.3ex\hbox{$<$\kern-0.75em\raise-1.1ex\hbox{$\sim$}}}
\def\gsim{\raise0.3ex\hbox{$>$\kern-0.75em\raise-1.1ex\hbox{$\sim$}}}

\begin{document}

\title{DIS AT LOW X, PHENOMENOLOGICAL ASPECTS
\thanks{Supported by DFG, contract Schi 189/6-2}}

\author{D. SCHILDKNECHT \\
Fakult{\"a}t f{\"u}r Physik, Universit{\"a}t Bielefeld \\
Universit{\"a}tsstrasse 25, 33615 Bielefeld, Germany\\
E-mail: Dieter.Schildknecht@physik.uni-bielefeld.de }

\maketitle

\begin{abstract}
\noindent Saturation at low x appears as an almost unavoidable consequence of
the two-gluon exchange generic structure. Consistency of the ansatz for the
vector part of the color dipole cross section with conventional evolution
determines the energy dependence of the saturation scale.
\end{abstract}



\markboth{\large \sl D. Schildknecht  \hspace*{2cm} HSQCD 2004}
{\large \sl \hspace*{1cm} DIS AT LOW X, PHENOMENOLOGICAL ASPECTS}

In this written version of my talk I will restrict myself to a discussion
of the empirical evidence for the concept of ``saturation'' at low x in
deep inelastic lepton-nucleon scattering (DIS) and of a consistency
argument that allows one to predict the energy dependence  of the
saturation scale.

In the model-independent analysis of the experimental data from HERA on DIS
at low x carried out in the summer of the year 2000, 
we found\cite{Diff2000}
that the data
on the total virtual photoabsorption cross section lie on a universal curve
when plotted against the dimensionless variable
\vspace*{-0.2cm}
\begin{equation}
\eta = \frac{Q^2 + m^2_0}{\Lambda^2_{sat} (W^2)},\label{(1)}
\end{equation}\vspace*{-0.2cm}
where\vspace*{-0.2cm}
\begin{equation}
\Lambda^2_{sat} (W^2) = B \left( \frac{W^2}{W^2_0} + 1 \right)^{C_2} \simeq B
\left( \frac{W^2}{W^2_0} \right)^{C_2}.\label{(2)}
\end{equation}
Compare fig. 1. The energy-dependent quantity, $\Lambda^2_{sat} (W^2)$, acts
as the scale (``saturation scale'' or ``saturation momentum'') that determines
the range of $Q^2$ in which the energy dependence (at fixed $Q^2$) is
either hard $(\eta >> 1)$ or soft $(\eta << 1)$. The model-independent analysis
only rest on the assumption that $\sigma_{\gamma^*p} (W^2, Q^2)$ be a smooth
function of $\eta$. The fitting procedure gave\cite{Diff2000,Cvetic}
\vspace*{-0.1cm}
\begin{eqnarray}
m^2_0 & = & 0.15 \pm 0.04 GeV^2,\nonumber \\
W^2_0 & = & 1081 \pm 12 GeV^2, \nonumber \\
C_2 & = & 0.27 \pm 0.01.\label{(3)}
\end{eqnarray}
As long as only smoothness of $\sigma_{\gamma^*p}$ is assumed, the constant
$B$ can be arbitrary. With the explicit form of $\sigma_{\gamma^*p}$ in the
generalized vector dominance-color dipole picture (GVD-CDP), we found
\vspace*{-0.1cm}
\begin{equation}
B = 2.24 \pm 0.43 GeV^2.\label{(4)}
\end{equation}

\begin{figure}[!thb]
\vspace*{4cm}
\begin{center}
\includegraphics{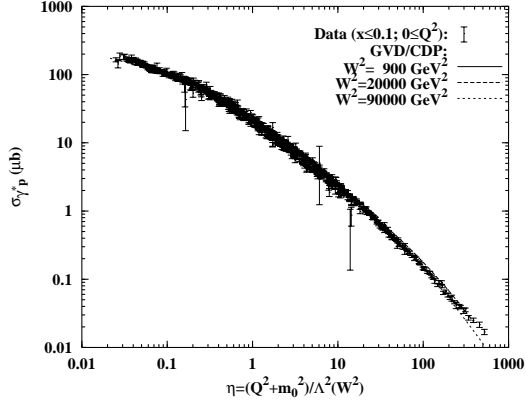}
\vspace*{1.5cm}
\caption[*]{The total photoabsorption cross section as a function of the 
scaling variable $\eta$ from (1). }
\end{center}
\label{fig1}
\end{figure}

\begin{figure}[!thb]
\vspace*{5.0cm}
\begin{center}
\includegraphics{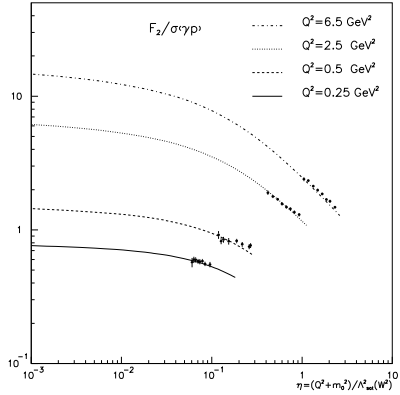}
\caption{The ratio of the structure function $F_2 (x, Q^2)$ and the
photoabsorption cross section as a function of $\eta$.}
\end{center}
\label{fig2}
\end{figure}
Note that the data shown in fig. 1 include all data available for $x \simeq
Q^2/W^2 < 0.1$ and $0 \le Q^2 < 1000 GeV^2$, in particular, photoproduction
$(Q^2 = 0)$ is included.

Since the HERA energy, $W$, is limited, for large values of $Q^2$ small values
of $\eta << 1$ cannot be explored. The low-$\eta$ region in fig. 1 contains
data close to photoproduction, while the large-$\eta$ region is populated by
large-$Q^2$ measurements. Nevertheless, fig. 1 suggests that the ``saturation''
property\cite{Diff2000,Cvetic}
\vspace*{-0.1cm}
\begin{equation}
\lim_{{W^2 \to \infty}\atop{Q^2 fixed}} \frac{\sigma_{\gamma^*p} 
(\eta (W^2, Q^2))}{\sigma_{\gamma p} (W^2)} = 1\label{(5)}
\end{equation}
to be valid for any fixed $Q^2$.

In terms of the structure function
\vspace*{-0.1cm}
\begin{equation}
F_2 (x, Q^2) \simeq \frac{Q^2}{4 \pi^2 \alpha} \sigma_{\gamma^*p} (\eta
(W^2, Q^2)),\label{(6)}
\end{equation}
where $x \simeq Q^2/W^2$, according to (5) we have
\vspace*{-0.1cm}
\begin{equation}
\lim_{{W^2 \to \infty}\atop{Q^2 fixed}} 4 \pi^2 \alpha 
\frac{F_2 (x, Q^2)}{\sigma_{\gamma p}(W^2)} = Q^2.\label{(7)}
\end{equation}
An explicit empirical test of the approach to saturation accordingly requires
to plot the data for the ratio of the structure function $F_2 (x, Q^2)$ and
the photoproduction cross section as a function of $\eta$ at fixed $Q^2$.
Saturation requires the ratio (7) to become flat and approach the value
of $Q^2$ as a function of $\eta$ as soon as $\eta$ becomes small,
$\eta << 1$. 

The plot of the experimental data in fig. 2\cite{Schi-Ku}, 
for $Q^2 \lsim 0.5 GeV^2$ shows
the expected flattening in the $\eta$-dependence for $\eta << 1$. For larger
values of $Q^2$ the expected flattening for $\eta \lsim 0.1$ cannot be 
verified at present due to lack of energy.

No explicit theoretical ansatz is needed for the plots in figs. 1 and 2.
We have nevertheless included the theoretical curves 
from the GVD-CDP\cite{Diff2000,Cvetic,Ku-Schi} that
provides a theoretical basis for the observed scaling in 
$\eta$.

As conjectured\cite{Sakurai,Fraas} a long time ago, 
DIS at low x in terms of the 
virtual-photon-proton Compton amplitude is to be understood in terms of 
diffractive forward scattering of the hadronic $(q \bar q)^{J=1}$ (vector) 
states the virtual photon dissociates or fluctuates into. With the advent
of QCD, the underlying Pomeron exchange became understood in terms of
the coupling of two gluons\cite{Low} to the $(q \bar q)^{J=1}$ state. The
gauge-theory structure implies that the $(q \bar q)^{J=1}_{T,L}~p$
color-dipole cross section, proportional to the imaginary part of the 
$(q \bar q)^{J=1}_{T,L}~p$ forward-scattering amplitude, 
takes the form\cite{Nikolaev,Ku-Schi}
\vspace*{-0.2cm}
\begin{eqnarray}
&&\hspace*{-0.8cm} \sigma_{(q \bar q)^{J=1}_{T,L} p} 
(\vec r_\perp^{~\prime}, W^2)  \int
d^2 \vec l^{~\prime}_\perp \bar \sigma_{(q \bar q)^{J=1}_{T,L}p} 
(\vec l^{~\prime 2}_\perp, W^2) \cdot 
(1 - e^{-i\vec l^{~\prime}_\perp \vec r^{~\prime}_\perp})\\
& \simeq & \sigma^{(\infty)} \begin{cases}
1, & $$ {\rm for}~\vec r^{~\prime 2}_\perp \to \infty,$$ \nonumber \\
\frac{1}{4} \vec r^{~\prime 2}_\perp \Lambda^2_{sat} (W^2), & $${\rm for}~
\vec r^{~\prime 2}_\perp \to 0,$$
\end{cases} \label{(8)}
\end{eqnarray}
where by definition
\begin{equation}
\sigma^{(\infty)} \equiv \pi \int d \vec l^{~\prime 2}_\perp 
\bar \sigma_{(q \bar q)^{J=1}_L} (\vec l^{~\prime 2}_\perp, W^2), \label{(9)}
\end{equation}
and 
\vspace*{-0.2cm}
\begin{equation}
\Lambda^2_{sat} (W^2) \equiv \frac{\pi}{\sigma^{(\infty)}} 
\int d l^{~\prime 2}_\perp \vec l^{~\prime 2}_\perp 
\bar \sigma_{(q \bar q)^{J=1}_L} (\vec l^{~\prime 2}_\perp, W^2). \label{(10)}
\end{equation}
The (virtual) photoabsorption cross section is obtained from (8) by 
multiplication with the (light-cone) photon-wave function and subsequent
integration over the transverse  $q \bar q$ separation 
$\vec r_\perp = \vec r_\perp^{~\prime}/ \sqrt{z (1-z)}$ and the variable 
$z$ with $0 \le z
\le 1$ that e.g. determines angular distribution of the quark in the
$q \bar q$ rest frame.

It is important to note that the two-gluon-exchange dynamical mechanism 
evaluated for $x \to o$ implies the existence  of the saturation scale
$\Lambda^2_{sat} (W^2)$ according to (8). The scale $\Lambda^2_{sat}
(W^2)$ is related to the effective value of the gluon transverse momentum,
$\vec l_\perp = \vec l^{~\prime}_\perp \sqrt{z(1-z)}$, that enters the 
photoabsorption cross section as a consequence of the two-gluon-exchange
mechanism. While an energy-independent scale $\Lambda^2_{sat} = {\rm const}$
a priori cannot be strictly excluded, it appears theoretically unlikely. Among
other things, constancy would mean that the effective gluon transverse
momentum from (10) would be energy independent, the diffractively
produced $q \bar q$ mass spectrum be energy independent, the full $W$
dependence reduced to a factorizing $W$ dependence due to 
a potential (weak) energy dependence of $\sigma^{(\infty)}$
alone, etc. The generic two-gluon-exchange structure ``almost'' rules
out $\Lambda^2_{sat} = {\rm const}$ and accordingly requires saturation.

Taking advantage of the fact that the ($J = 1$ part of the) dipole
cross section (8) is essentially determined by the quantities 
$\sigma^{(\infty)}$ and $\Lambda^2_{sat} (W^2)$ in 
(9) and (10), the
total photoabsorption cross section becomes 
approximately\cite{Diff2000,Cvetic,Ku-Schi}
\begin{equation}
 \sigma_{\gamma^*p} (W^2, Q^2) \simeq 
\frac{\alpha R_{e^+e^-}}{3 \pi} \sigma^{(\infty)} 
\begin{cases}
ln \frac{\Lambda^2_{sat} (W^2)}{Q^2 + m^2_0}, 
& $$(Q^2 << \Lambda^2_{sat} (W^2)),$$\\
\frac{1}{2} \frac{\Lambda^2_{sat} (W^2)}{Q^2} 
& $$ (Q^2 >> \Lambda^2_{sat} (W^2)).$$
\end{cases}
\label{(11)}
\end{equation}
A detailed evaluation leads to the theoretical results displayed in figs.
1 and 2.

Several remarks are appropriate:
\begin{itemize}
\item[i)] Unitarity for the hadronic $(q \bar q)$ proton interaction
requires the integral (9) to exist and $\sigma^{(\infty)}$ to be
at most weakly dependent on the energy $W$. The fit yields
$\sigma^{(\infty)} \simeq {\rm const}~ \simeq 30 mb.$
\item[ii)] The existence of a scale, $\Lambda^2_{sat} (W^2)$, 
according to (10) appears as a 
straightforward consequence of the two-gluon-exchange structure. This structure
implies that the forward-scattering amplitude depends on the effective gluon
transverse momentum.
\item[iii)] Since unitarity $(\sigma^{(\infty)} \simeq {\rm const})$
cannot be disputed, and the assumed two-gluon exchange generic structure
seems safe, once $\Lambda^2_{sat} = {\rm const}$ is abandoned, we must
have the transition to the logarithmic behavior in (11), i.e. saturation
as depicted in fig. 2 even far beyond the energy range accessible at
present.
\item[iv)] The gluon structure function from (8) is 
given by\cite{Zakharov} 
\begin{equation}
\alpha_s
(Q^2) x g (x, Q^2) = \frac{1}{8 \pi^2} \sigma^{(\infty)} \Lambda^2_{sat}
\left(\frac{Q^2}{x}\right),\label{(12)}
\end{equation}
again disfavoring constancy of $\Lambda^2_{sat}
(W^2)$.\\ 
\item[v)] When saturation and the logarithmic behavior in (11) set in, the
usual connection between $F_2$ and the gluon structure function breaks
down. An extensive literature (compare e.g.\cite{Iancu} and the references
therein) attempts to apply (nonlinear) evolution equations for gluon 
distributions even in this logarithmic domain. 
\end{itemize}

\begin{figure}[!thb]
\vspace*{10.0cm}
\begin{center}
\includegraphics{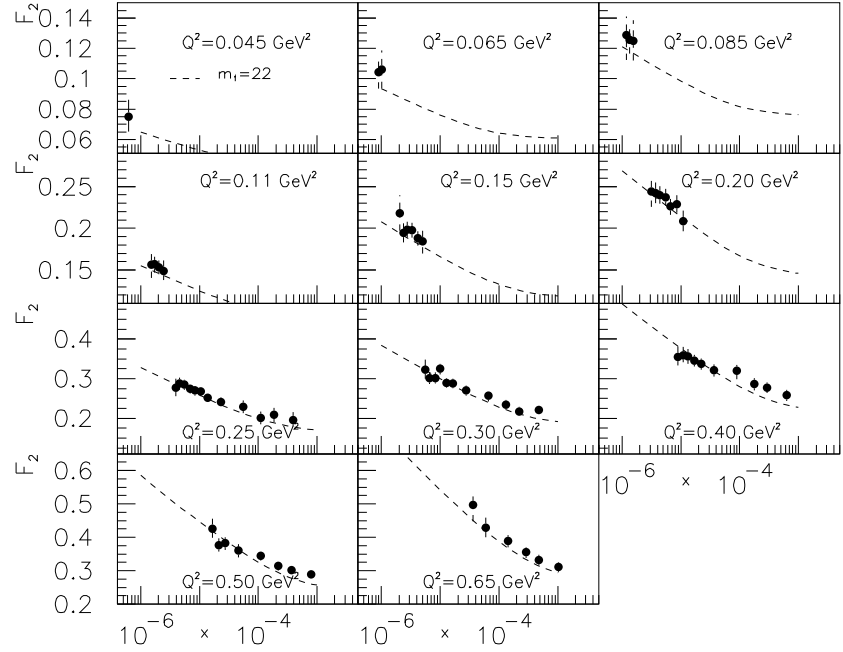}
\caption{The HERA data for the structure function $F_2 (x, Q^2)$ with the
predictions of the GVD-CDP.}
\end{center}
\label{fig2}
\end{figure}

\begin{figure}[!thb]
\vspace*{13cm}
\begin{center}
\includegraphics{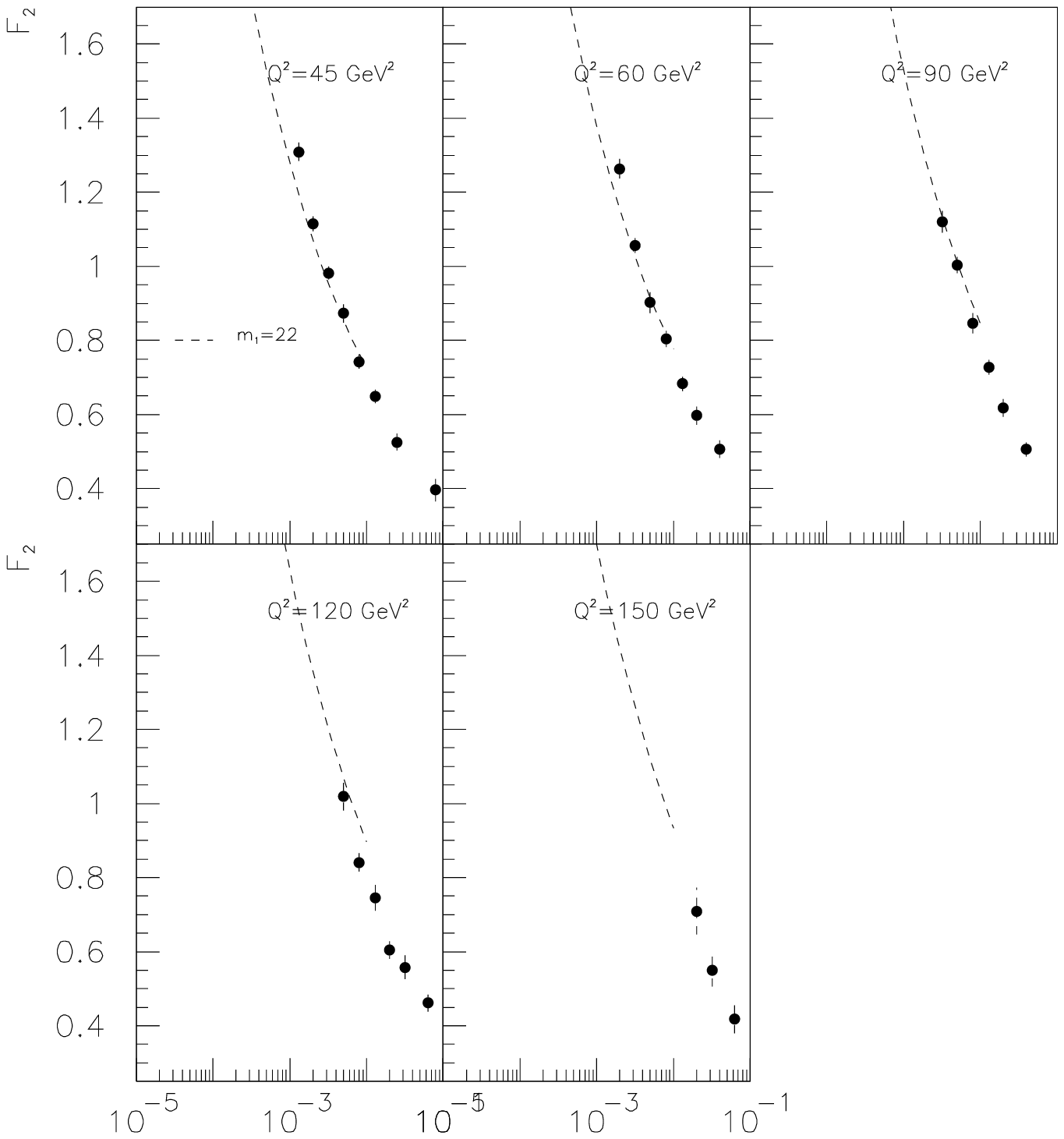}
\caption{Same as fig. 3 but for large values of $Q^2$.}
\end{center}
\label{fig2}
\end{figure}

We examine the theoretical description of the experimental data for 
$F_2 (x, Q^2)$ in somewhat more detail. The ansatz for the dipole cross
section underlying the results depicted in fig. 1 and fig. 2 is given by
(8) with\cite{Cvetic,Ku-Schi}
\begin{equation}
\bar \sigma_{(q \bar q)^{J=1}_T} (\vec l_\bot^{~\prime 2}, W^2) = 
\bar \sigma_{(q \bar q)^{J=1}_L} (\vec l_\bot^{~\prime 2}, W^2) =
\sigma^{(\infty)} \frac{1}{\pi} \delta (\vec l_\bot^{~\prime 2} -
\Lambda^2_{sat} (W^2)).\label{(13)}
\end{equation}
The total virtual photoabsorption cross section may be represented in
terms of the diffractive forward production of discrete and continuum
$(q \bar q)^{J=1}$ (vector) states\cite{Ku-Schi}. The upper limit in the 
mass of such states, $M^2_{(q \bar q)} \lsim m^2_1 \simeq 484 Gev^2$
at HERA energies, enters the description of $\sigma_{\gamma^*p} (\eta)$ at
large values of $\eta$. In fig. 3 and fig. 4, as an example, we show 
$F_2 (x, Q^2)$ for several values of $Q^2$.

In (12), we noted the connection between the dipole cross section (8) in the 
limit of small interquark separation and the gluon structure function of
the proton. At low x and sufficiently large values of $Q^2$, the change of
the structure function $F_2(x, Q^2)$ is determined by the gluon structure
function alone via\cite{Prytz}
\begin{equation}
\frac{\partial F_2 (\frac{x}{2}, Q^2)}{\partial ln Q^2} =
\frac{R_{e^+e^-}}{9 \pi} \alpha_s (Q^2) xg (x, Q^2),
\label{(14)}
\end{equation}
where $R_{e^+e^-} = 3 \sum Q^2_f = 10/3$ for four flavors of quarks with
charges $Q_f$.
Substituting the expression for $F_2 (x,Q^2)$ at large $Q^2$ from (11)
and (6) on the left-hand side in (14), and the expression (12) for the
gluon structure function on the right-hand side, we obtain an interesting 
consistency constraint on the saturation scale that reads
\begin{equation}
\frac{\partial}{\partial ln W^2} \Lambda^2_{sat} (2W^2) = \frac{1}{3} 
\Lambda^2_{sat} (W^2).
\label{(15)}
\end{equation}
Substitution of the asymptotic power law (2) implies\cite{Cvetic,Schi-Ku}
\begin{equation}
C_2^{theory} = \frac{1}{3} \left(\frac{1}{2}\right)^{C_2}
\label{(16)}
\end{equation}
or
\begin{equation}
C_2^{theory} = 0.276.\label{(17)}
\end{equation}
Consistency with conventional DGLAP evolution of our ansatz (13) for
the $J=1$ (vector) part of the color dipole cross section implies the
prediction (17) for the energy dependence of the saturation scale (2). The
prediction (17) is consistent with the experimental value of $C_2$ in (3)
deduced from the analysis of the experimental data, where $C_2$ was left
as a free parameter. Note that this predicted energy dependence heavily
relies on the choice of $W$ as the relevant variable in the dipole
cross section and the saturation scale, $\Lambda^2_{sat} = \Lambda^2_{sat}
(W^2) = \Lambda^2_{sat} (Q^2/x)$. Moreover, it relies on the assumed 
equality of the scattering of longitudinally and transversely polarized
$(q \bar q)^{J=1}$ states (13), a constraint that is imposed beyond the
generic two-gluon-exchange structure.\footnote{I am indebted to Borilav
Zakharov for valuable discussions on the assumptions underlying the result
for $C_2$.} The available data on the longitudinal-to-transverse ratio
$\sigma_{\gamma^*_L p}/\sigma_{\gamma^*_T p}$ are consistent with this
constraint\cite{Tent}. Further investigation on this significant result
on the energy dependence of the saturation scale, e.g. its relation to
the double-log approximation of the gluon structure function are in
progress\cite{Schi-Ku}.

\section*{Acknowledgements}
It is a pleasure to thank Masaaki Kuroda for a fruitful collaboration
and Lev Lipatov and Victor Kim for a splendid organization and a 
wonderful hospitality in Repino and St. Petersburg.

\end{document}